\documentstyle[aaspp4]{article}

\def\marc{$\rm mag~arcsec^{-2}$}
\lefthead{Gavazzi et al.}
\righthead{Star Formation in disk galaxies}

\begin{document}
\title{The star formation properties of disk galaxies: $\rm H\alpha$ imaging of galaxies in the 
Coma supercluster\footnote{based on observations made at the Observatorio Astr\'onomico Nacional (OAN), San Pedro M\'artir, B.C. (OAN) of the Universidad Nacional  Aut\'onoma de Mexico}}

\author{Giuseppe Gavazzi\altaffilmark{2} and Barbara Catinella}
\affil{Universita' degli Studi di Milano, dipartimento di Fisica, via Celoria, 16, 20133 Milano, 
Italy\\
Electronic mail: gavazzi@brera.mi.astro.it, catinel@trane.uni.mi.astro.it}

\author{Luis Carrasco\altaffilmark{3}}
\affil{Instituto Nacional de Astrof{\'i}sica, Optica y Electr\'onica,
Apartado Postal 51. C.P. 72000 Puebla, Pue., M\'exico \\
Electronic mail: carrasco@inaoep.mx}

\author{Alessandro Boselli}
\affil{Laboratoire d' Astronomie Spatiale BP8, Traverse du Syphon, F-13376 Marseille, France\\
Electronic mail: boselli@astrsp-mrs.fr}

\author{Alessandra Contursi}
\affil{DEMIRM, Observatoire de Paris, 61 Av. de l'Observatoire, F-75014 Paris, France\\
Electronic mail: contursi@mesioq.obspm.fr}

\altaffiltext{2}{Osservatorio Astronomico di Brera, via Brera 28, 20121, Milano, Italy}

\altaffiltext{3}{Also OAN/UNAM, Ensenada B.C., M\'exico}


\begin{abstract}

We present integrated $\rm H\alpha$ measurements obtained from imaging observations of 98 late-type  
galaxies, primarily selected in the Coma 
supercluster. 
These data, combined with $\rm H\alpha$ photometry from the literature, 
include a magnitude selected sample of spiral (Sa to Irr) galaxies belonging to the ``Great Wall'' complete up to 
$\rm m_p=15.4$, thus composed of galaxies brighter than $\rm M_p=-18.8$ 
($\rm H_0=100~km~Mpc^{-1}~s^{-1}$). The frequency distribution of the $\rm H\alpha$ E.W., determined for 
the first time from an optically complete sample,  is approximately gaussian peaking at E.W.$\rm \sim~25~\AA$. 
We find that, at the present limiting luminosity, the star formation properties of spiral+Irr galaxies 
members of the Coma and A1367 clusters do not differ significantly from those of the isolated ones belonging to the Great Wall. 

The present analysis confirms the well known increase of the current massive star formation rate (SFR) with Hubble type. Moreover perhaps a more fundamental anticorrelation exists between
the SFR and the mass of disk galaxies: low-mass spirals and dwarf systems have present SFRs 
$\sim$ 50 times higher than giant spirals. 
This result is consistent with the idea that disk galaxies are coeval, evolve as ``closed systems'' with exponentially declining SFR
and that the mass of 
their progenitor protogalaxies is the principal parameter governing their evolution. 
Massive systems having 
high initial efficiency of collapse, or a 
short collapse time-scale, have retained little gas to feed the present epoch of star formation.
These findings support the conclusions of Gavazzi \& Scodeggio (1996) who studyed the 
color-mass relation of a local galaxy sample and agree with the analysis by Cowie et al. (1996) who traced the star formation history of galaxies up to z$>$1.
\end{abstract}



\section{Introduction}

Two major aspects of the phenomenology of disk galaxies, with relevant implications on 
the theory of galaxy formation and evolution, are still hotly debated among the scientific 
community: \nl
(i) whether the color-luminosity relation, well established among late-type 
galaxies, follows from a population sequence, as proposed by Gavazzi, Pierini \& Boselli
(1996) (hereafter GPB96) and by Gavazzi \& Scodeggio (1996) (hereafter GS96), as opposite 
to a metallicity sequence (see Visvanathan 1991; Bothun et al. 1984), which appears to be the case in 
elliptical galaxies (see Arimoto \& Kodama, 1997); \nl
(ii)  whether the current star formation rate (SFR) of spiral 
galaxies belonging to a rich cluster is significantly lower than that of isolated galaxies of similar 
morphological type, as expected if either formation or evolutionary processes (e.g. ram-pressure) 
would contribute depleting the gaseous content of spiral galaxies in clusters,
thus reducing the gaseous reservoir necessary to feed their star formation.

Kennicutt's pioneering work in this field helped to establish that the massive ($\rm >10~M_{\sun}$), 
current ($\rm t<10^7~yrs$) star formation rate of disk galaxies is accurately traced by the 
integrated $\rm H\alpha$ + [NII] line intensity (Kennicutt 1983a) normalized to the underlying 
continuum intensity (Kennicutt 1989, 1990), i.e. by the line equivalent width (E.W.). 
More recently Kennicutt, Tamblyn \& 
Congdon (1994, hereafter KTC94) determined that the ratio of the present to past rate 
increases from 0.01 to 1 
along the Hubble sequence (from Sa to Irr), reflecting a change in the star formation 
properties of disks, and only secondarily a change in the bulge-to-disk ratio.

To investigate the hypothesis i), in Section 5.1 of this paper we study the 
dependence of the $\rm H\alpha$ E.W. on near-infrared H band luminosity. Our large 
(although not-complete) sample of galaxies spans a broad range of luminosities for which 
$\rm H\alpha$ E.W. and H band photometry are available. Following GPB96 we assume that the H 
band luminosity is proportional to the galaxy dynamical mass.

To discuss point ii), we compare the systematic (i.e. derived from a survey having the 
character of completeness) $\rm H\alpha$ properties of spiral galaxies as a function of some 
indicator of their environmental conditions, e.g. their local galaxy density or the projected 
radial distance from the cluster centers. \nl
To this purpose Kennicutt \& Kent (1983, KK83 hereafter) carried out an $\rm H\alpha$ survey of 
galaxies belonging to the Virgo cluster and compared them with isolated objects. Kennicutt (1983b) 
tentatively concluded that galaxies belonging to the core of this cluster have a  SFR significantly 
quenched in comparison to normal galaxies. Kennicutt, Bothun \& Schommer (1984, 
KBS84 hereafter) extended this study to other dynamical entities, such as the Cancer cluster, Coma 
and A1367. The sample they collected, however, was not sufficient to derive general 
conclusions. Neither sufficiently complete was the $\rm H\alpha$ study by Gavazzi, Boselli \& 
Kennicutt (1991, GBK91 hereafter) who carried out an aperture photometry, narrow band survey 
of another 55 late-type galaxies in the Coma ridge. They found that a significant number of 
S+Irr galaxies projected onto the two clusters have surprisingly strong $\rm H\alpha$ emission.
These include the blue galaxies found in the Coma cluster by Bothun \& Dressler (1986). \nl
A problem affecting these early studies was the lack of a well defined zero point for their 
environmental comparison, i.e. the properties of isolated galaxies were not known with 
sufficient accuracy. With the aim of establishing the present high mass  SFR of spiral 
galaxies in the Universe and of comparing these properties in environments that differ by 
an order of magnitude in local galaxy density we undertook the present $\rm H\alpha$ imaging 
survey. The Coma supercluster offers a unique test-bed for such systematic study since it 
contains a filament of galaxies at constant distance from us (the ``Great Wall''), thus a 
volume limited sample can be extracted from a magnitude limited sample. Moreover the 
properties of galaxies in two rich clusters such as Coma and A1367 (with a density of 
about 3 galaxies $\rm Mpc^{-3}$ brighter than 15.7) can be compared with those of relatively 
isolated galaxies (environments with about 10\% of the cluster density), lying in the bridge between the two 
clusters at a similar distance from us, thus suffering from similar distance biases. 
 
\section{The Sample}

Imaging in the light of the $\rm H\alpha$+[NII] line of galaxies selected primarily in the region of 
the Coma supercluster is reported in this work. Out of the 262 spiral galaxies brighter than 
$\rm m_p=15.7$ belonging to the ``Great Wall'', listed in the catalogue of Gavazzi \& Boselli (1996), 
we observed 90 objects (8 additional early-type galaxies were observed serendipitously
in the target fields. These will be not used in the following analysis). 
By themselves these objects do not constitute a complete sample. 
However, from the observations presented in this paper in conjunction with the $\rm H\alpha$ 
aperture photometry measurements taken from the literature (KK83, KBS84, Moss, 
Whittle \& Irwin 1988 (MWI88), Romanishin 1990, GBK91), two complete subsamples can 
be extracted:

\noindent
the ``cluster'' sample contains 40 (28 imaged in this work +12 from the literature) 
spiral galaxies with $\rm m_p\leq15.4$, belonging to either A1367 or to the Coma cluster 
(A1656);

\noindent
the ``isolated'' sample contains 66 (44 imaged  in this work + 22 from the 
literature) galaxies with $\rm m_p\leq15.4$ in the ``Great wall'' defined as isolated, 
i.e. those whose nearest companion lies at a projected distance of more than 300 kpc. 

\noindent
These two samples are used in the present investigation to derive frequency distributions 
which need to be extracted from a magnitude complete sample (Sects. 5.3-5.4). The remaining 
objects observed in this work and/or found in the literature will be used only in those 
analysis which do not depend upon completeness (Sect. 5.1, 5.2).
We remark that, since the studied galaxies belong to a structure (the Coma supercluster) 
approximately at constant distance from the observer,
a volume limited sample can be easily extracted from the present magnitude
complete sample.

\noindent
Table 1 lists the observed galaxies as follows:

\noindent
Col. 1: CGCG designation (Zwicky et al. 1961-1968). \nl
Col. 2-3: NGC/IC, UGC (Nilson 1973) names. \nl
Col. 4 and 5: 1950 celestial coordinates (1-2 arcsec uncertainty). \nl
Col. 6 and 7: major and minor axes (in arcmin) as determined at the 25 \marc 
(see Gavazzi \& Boselli 1996). \nl
Col. 8: heliocentric velocity from the literature. \nl
Col. 9: cluster membership (see Gavazzi \& Boselli 1996 for more details).
Members of Coma and A1367 are those objects with a projected angular separation within 
2 and 1 degrees respectively from their X-ray centers. One galaxy belongs to the N4794 
group, another to the N3937 group in the foreground of the Coma supercluster. 
Two objects constitute a pair of galaxies in the Coma supercluster. 
Six filler galaxies belong to the Cancer cluster, two
are in the foreground of the 
Hercules supercluster and one is a member of A2197. 
The remaining galaxies are isolated supercluster objects.
We assume a distance of 65 Mpc 
for A1367, 69 for Coma. Isolated supercluster objects are assumed at their redshift 
distance, using $\rm H_0=100~km~Mpc^{-1}~s^{-1}$. \nl
Col. 10: the cluster membership according to the less restrictive criterion of Gavazzi, 
Randone \& Branchini (1995). A galaxy is considered a cluster member if it lies within the ``caustic'' associated with the cluster potential well. \nl
Col. 11: projected radial distance from the nearest cluster center (deg). \nl
Col. 12: morphological type (see Gavazzi \& Boselli 1996). The classification was performed on the
best photographic material available: a) the present frames; b) broad band CCD frames; c)
KPNO 4m plates (for the two clusters); d) Palomar Sky Survey plates.
The majority of galaxies in our sample have small angular size (1-2 arcmin), thus the
classification error is probably up to 2 bins of Hubble type. In particular the overabundance
of Pec and Irr objects reflects our classification inability.
Given the distance of the Coma supercluster no Magellanic Irr type is included in this work.
We classify Irr any galaxy with no evident spiral structure nor circular symmetry.
We prefer to classify Pec the otherwise normal objects with superposed peculiar features, such
as bright spots, extra arms etc. Pec/c stands for Peculiar/compact; Pec/ms for Peculiar/multiple
system. \nl 
Col. 13: photographic magnitude from Zwicky et al. (1961-1968). \nl
Col. 14: $\rm B_0^T$ magnitude corrected for internal extinction (see Gavazzi \& Boselli 1996). \nl
Col. 15: $\rm H_0^T$ magnitude corrected for internal extinction (see Gavazzi \& Boselli 1996).

\noindent
The sky distribution of CGCG galaxies members to the Coma Supercluster, thus restricted 
to the velocity range 5000-10000 $\rm km~s^{-1}$, is given in Fig. 1. 

\section{Observations and Data Reduction}

Narrow band imaging of the $\rm H\alpha$ line emission ($\rm \lambda=6562.8~ \AA$) from the target galaxies 
was obtained during from 1993 to 1997, using the 2.1 m telescope at San Pedro 
Martir Observatory (SPM) (Baja California, Mexico). Four galaxies were observed in 1990 using the 
Steward 2.3 m telescope at Kitt Peak (STW) (see also Gavazzi et al. 1995). The STW 
telescope was equipped with a 380x400 pixel CCD coupled with a focal reducer with a 
resulting pixel size of 1.2 arcsec. 
The SPM Cassegrain focus (f/7.5) was coupled with a 1024x1024 pixel Thx31156 CCD in 1993 
and 1994, with 0.25 arcsec/pixel and a gain of 2.12 $\rm e^-$/ADU. Since 1995 the system was 
upgraded with a 1024x1024 pixel TEK1024AB CCD with nearly double quantum 
efficiency, 0.30 arcsec/pixel and a gain of 4.88 $\rm e^-$/ADU.

Each galaxy was observed using two narrow band interferometric filters: one that includes 
the redshifted wavelength of the $\rm H\alpha$ line (on), the other, of similar bandwidth, 
centered at least 100 $\rm \AA$ off the line (off) to secure the continuum measurement. The flux from the 
[NII] emission lines ($\rm \lambda$ 6548 $\rm \AA$ and 6584 $\rm \AA$) is included in the on-band observations.
Fig. 2 shows the transmission profiles of the filters used (as provided by the 
manufacturer for $\rm T=20^o$ C). Fig. 2a refers to the 1990-1995 data. The 
transmission curves of the SPM filters were remeasured in 1996 and are presented in Fig. 2b. These show a slightly lower transmission 
than previously measured. We allowed for a transmission shift toward the blue of 1 $\rm \AA$ per 5 deg of 
temperature decrease. Since we observed with a dome temperature around $\rm 10^o$ C, we 
applied a 2 $\rm \AA$ correction. At the redshift of the target galaxies the $\rm H\alpha$ line lies well 
within the region of maximum filter transmission.

\noindent
Table 2 reports the journal of observations as follows:

\noindent
Col. 1: CGCG galaxy name. \nl
Col. 2, 3: central wavelength of the filters used, on-band and off-band respectively. \nl
Col. 4: FWHM of the bandpass. \nl
Col. 5: telescope used. \nl
Col. 6: pixel scale in arcsec. \nl
Col. 7: observing date (dd-mm-yy). \nl
Col. 8: integration time per filter (minutes). \nl
Col. 9: flag indicating if the frame was taken under photometric conditions. Galaxies 
marked ``N'' were observed in nearly photometric conditions. \nl
Col. 10: normalization factor obtained dividing the flux of several field stars in the off-band frames by the 
flux in the on-band frames. In photometric conditions this quantity reflects the transmission difference between 
the on and off band filters and ranges between 0.95 and 1.20 (for the various filters and years). Under non-photometric conditions the quantity includes variations in the sky transparency. 

\noindent
All images were obtained in seeing conditions in the range 1.3-2.5 arcsec.
 
\subsection{Image Analysis}

The data reduction of the present frames was based on the IRAF package, 
developed by NOAO, and on the SAOIMAGE package, developed at the Center for 
Astrophysics. To remove the detector response every raw image was bias subtracted and 
divided by the flat-field obtained from the twilight sky. 
Cosmic rays were individually removed from each frame.
Bad pixel columns were removed by direct inspection of the frames and 
replaced with the mean of adjacent columns.
The sky background was determined in each frame in concentric object-free annuli about 
the object and then subtracted from the flat-fielded images. These resulting frames were then used to determine the total counts from the objects of interest (galaxies and stars). \nl
Let us consider a frame containing a target galaxy (with total counts $cnts^g$ ) and one or 
more field stars (with total counts $cnts^*$). Under the assumption that the field stars do not 
emit $\rm H\alpha$, i.e. that: \nl
$flux^*_{on}~/~flux^*_{off}~=~1$ \nl
we have: \nl
$cnts^*_{on}~/~cnts^*_{off} = T_{on}\times \tau_{on}\times c~/~T_{off}\times \tau_{off}\times c = K$, \nl
where: \nl
$cnts~=~flux~\times T\times \tau \times c$ \nl (T = effective integration time;
$\rm \tau$= integral of the filter transmission; 
c= conversion factor from $cnts~s^{-1}$ to $erg~cm^{-2}~s^{-1}$).

\noindent
For a target galaxy the  $\rm H\alpha$  equivalent width (E.W.) is then:

\vskip 1truemm
\noindent
$E.W.~=~{flux^g_{on}~-~flux^g_{off}\over flux^g_{off}}$

\vskip 1truemm
$=~{cnts^g_{on}~-~K~cnts^g_{off}\over K~cnts^g_{off}}~~~~~[\AA]$

\vskip 1truemm
\noindent
independent from c, thus derivable also in non photometric conditions from the observed quantity K. \nl
The net flux in the  $\rm H\alpha$  line is given by:

\vskip 1truemm
\noindent
$net~flux~=~{cnts^g_{on}\over T_{on}~\tau_{on}~c}-{cnts^g_{off}\over T_{off}~\tau_{off}~c}~=~
{cnts^g_{on}~-~K~cnts^g_{off}\over T_{on}~\tau_{on}~c}~~~~~[erg~cm^{-2}~s^{-1}]$
\vskip 1truemm

\noindent
that requires a determination of c from the calibration process. \nl
In order to calibrate our data, we have observed the spectrophotometric stars Feige34, Hz44 and 
BD332642 just before or after the target galaxies. Their measured spectral energy distributions (available in 
IRAF in tabular form) were then convolved with the filter transmission profiles to obtain the value of c.

\section{Results}

Fig. 3a gives a gray-scale representation of the off (top panels) and net (on-off, bottom 
panels) frames of all galaxies with a net flux. Fig. 3b carries the off-frames of 11 galaxies with null or 
negative net flux.%
\footnote {The gif files containing figures 3a to 3o (gray-scale images) attached to the present version of the paper are of poorer quality than the original postscript files (approximately of 1.5 Mbytes size each). These are available upon request to G. Gavazzi (gavazzi@brera.mi.astro.it)}

\noindent
Table 3 reports the results of the present work as follows:

\noindent
Col. 1: CGCG galaxy name. \nl
Col. 2: telescope used. \nl
Col. 3: total equivalent width ($\rm \AA$) from the present work, jointly with its statistical uncertainty. \nl
Col. 4: logarithm of the integrated flux from the present work ($\rm erg~cm^{-2}~s^{-1}$). \nl
Col. 5: logarithm of the integrated luminosity ($\rm erg~s^{-1}$). The last two quantities are given only for 
targets observed under photometric conditions. \nl 
Col. 6, 7: equivalent width, with error and flux from the literature. \nl
Col. 8: reference.

\noindent
Fig. 4 shows the comparison between the E.W. (a) and flux (b) measurements reported in this paper and those found in the literature for the common objects. Fluxes and E.W. from KK83, KBS84 and 
GBK91 have been multiplied by 1.16, as suggested by KTC94, in order to account for the 
continuum flux overestimate due to the inclusion of the telluric absorption band near 6900 
$\rm \AA$ in the side-band filter.

\subsection{Comments to Individual Objects}

\noindent
97073: Fig. 3a reports the SPM net frame.

\noindent
97093: there is a severe disagreement for this galaxy between the present measurement 
and the value found by MWI88 in their objective prism survey of A1367. There are no apparent 
reasons to suspect  that our data, which were obtained under photometric sky conditions, are bad. Moreover, 
two early type galaxies (97088 and 97094) in the frame of 97093 (the redshift of 97094 
do not match the on-band filter) result, as expected, in a null net $\rm H\alpha$  flux. However, a measured U-B = -0.31 is 
consistent with the MWI88 $\rm H\alpha$ value.

\noindent
97129: stray light from a nearby bright star contaminates the frame.

\noindent
98078: this galaxy, companion to 98081, was classified as elliptical on the PSS, due
to its featureless appearence. However it is one of the strongest $\rm H\alpha$ emitters,
with 82 $\rm \AA$. This emission, almost entirely nuclear, is probably caused by the gravitational
interaction with its companion, and/or with a third fainter object in the vicinity 
(clearly seen in the net frame) whose redshift is not presently available. 

\noindent
160050: at the time of the observation presented 
in this paper (April 1995) a redshift of 5319 $\rm km~s^{-1}$ was available (Tifft \& Gregory 1976). 
Thus we used the 6683 $\rm \AA$ filter to get the on-band and the 6603 $\rm \AA$ for the off-band 
frame. After reducing the data we found, to our disappointment, that more flux and more 
structure showed up in the off-band frame. However, one year later we obtained a spectrum of this galaxy for 
another project. To our surprise we found that the true redshift is 2496 $\rm km~s^{-1}$. Thus it 
turns out that our $\rm H\alpha$ measurement is useful but with reversed filters.

\noindent
160055: the off-band frame was contaminated by some stray-light from a nearby bright 
star which produces a faint diffuse feature at the SE edge of the galaxy. The net flux from 
the galaxy might be slightly underestimated accordingly.

\noindent
224004S: no redshift is available for this object, thus the filters adopted for 224004 could
not match the $\rm H\alpha$ line. The $\rm H\alpha$ flux should be considered a lower limit.

\noindent
224038: this peculiar object in A2197 might be the remnant of a galaxy collision. The 
complex velocity field in this galaxy was studied by Maehara et al. (1988).

\section{Analysis}

To complement our imaging survey, we use in the following analysis $\rm H\alpha$ data taken
from the literature. 
Table 4 summarizes the data in the Cancer 
and Virgo clusters and in the Coma supercluster regions.

\noindent
Col. 1: CGCG designation (Zwicky et al. 1961-1968). \nl
Col. 2-3: NGC/IC, UGC names. \nl
Col. 4, 5: $\rm H\alpha$ equivalent width, along with its error and  source reference.

\subsection{The Dependence of the SFR on Hubble Type and Mass}

In this section we use the $\rm H\alpha$ E.W. reported in Tables 3 and 4 in order to study the relationship between the 
current star formation and other photometric properties of disk galaxies. 
First we show in 
Fig. 5a the relation between the  $\rm H\alpha$ E.W. and the morphological type.
The Hubble classification is represented by discrete numerical classes: 3=Sa, 4=Sab ... 7=Sc, 8=Irr/Pec. To avoid 
superposition of points, we added to each class a random number between -0.3 and 0.3.
The majority of galaxies in our sample have small angular size (1-2 arcmin), thus the
classification error is probably up to 2 bin in Hubble type. In particular, due to the
adopted classification scheme (see column 12 of Table 1) the Pec objects
might in fact belong to any of the earlier-type bins.
Nevertheless, in spite of the uncertainty on our classification, Fig. 5a shows a
definite trend of $\rm H\alpha$ E.W. with type. In fact the average $\rm H\alpha$ E.W.
among types Sa is 5.3 $\pm$ 1.9 $\AA$, 
significantly lower than 24.5 $\pm$ 3.3 $\AA$ found among types Sc (see also 
Roberts \& Haynes 1994 and KTC94).

To further investigate the reasons for the residual scatter found within each individual 
morphological class (see Fig. 5a), we analyze the dependence of the $\rm H\alpha$ E.W. 
on the galaxy's mass, a quantity carrying a more direct physical meaning than the Hubble type.
Following GPB96, we use the H band luminosity as a tracer of the dynamical 
mass of disk galaxies ($\rm Log~M_{dyn}=Log~L_H+0.66$, solar units) to show in
Fig. 5b that the $\rm H\alpha$ E.W. decreases significantly with increasing mass.
Although the relation appears noisy and non-linear, low mass systems ($\rm 9< Log~L_H < 10$) show high (40 $\rm \AA$) average E.W., weakly decreasing with mass. The one order of magnitude scatter found in this interval probably reflects
the presence or the absence of episods of star formation of short duration which is governed primarily
by the local gas instability (see Kennicutt, 1989). At ($\rm Log~L_H > 10$) the $\rm H\alpha$ E.W. drops to zero, with an even higher scatter than in low mass regime.
If one divides the data in 3 bins of decreasing $\rm H\alpha$ ($\rm 0<Log H\alpha < 1$; 
$1<Log H\alpha < 1.6$; $Log H\alpha > 1.6$) the corresponding average $\rm Log~L_H$ are: 10.9; 10.4; 10.0 $\rm L_{\sun}$, i.e. an approximately inverse linear proportionality.  

\noindent

\subsection{The Dependence of the SFR on the Bulge-to-Disk Ratio}

One of the most clear-cut results of GPB96 is that the presence in the Near-Infrared of centrally peaked structures 
(bulges-nuclei) strongly correlates with the total H (1.65 $\rm \mu$m) luminosity or mass
(see their Fig. 9). These authors used the model-independent parameter  $\rm C_{31}$,
defined as the ratio of the radii containing respectively 75\% and 25\% of the total H band light, 
and found $\rm C_{31}$  in the range 1-3 for pure disk objects, and  $\rm C_{31}~>~3$ in galaxies with 
prominent bulges. \nl
Given this result, it is important to investigate whether or not the inverse correlation between  SFR
and  mass, reported in the previous section, follows from the anti-correlation between $\rm H\alpha$ E.W. and 
$\rm C_{31}$. In other words, whether or not the $\rm H\alpha$ E.W. 
is strong in pure disk galaxies and weak in bulge dominated galaxies, as 
claimed by Devereux \& Young (1991). However KTC94 showed that variations of $\rm H\alpha$ E.W. do not simply 
reflect different contributions of bulge light to the continuum, but real variations of the disk SFR. \nl
Points in Fig. 5 a and b (and in the following figures, unless otherwise specified) are
coded according to $\rm C_{31}$: filled dots are pure disks, open symbols are bulge dominated 
systems and crosses indicate those objects for which no structural information is available.
Pure disks dominate the low $\rm L_H$ - high $\rm H\alpha$ - late-type regime.
Also in Fig. 6 (a, b, c), where we plot the $\rm H\alpha$ - $\rm L_H$ relation
in separate morphological type bins, the segregation according to $\rm C_{31}$ is apparent,
and enhanced in panels d and e.

It is evident that the quantities $\rm C_{31}$, $\rm H\alpha$, $\rm L_H$ and morphological type
are not independent of one another, as illustrated in Fig. 7.
However, while both $\rm H\alpha$ (Fig. 7a, extracted from Fig. 5b)
and $\rm C_{31}$ (Fig. 7b) are strongly related with $\rm L_H$, the correlation of  
$\rm H\alpha$ with $\rm C_{31}$ (Fig. 7c) is poorer:
most disk galaxies ($\rm C_{31}<$3) have high SFR, but a significant fraction of them
have  $\rm Log H\alpha<$1. Any value of $\rm H\alpha$ E.W. is found among bulge-dominated objects.

In conclusion, we claim that the dependence of $\rm H\alpha$ E.W. on mass (H luminosity) is
not entirely induced by the dependence of the bulge-to-disk ratio on luminosity. The primary dependences are 
those of $\rm C_{31}$ and of $\rm H\alpha$ E.W. on mass,
and the one of $\rm H\alpha$ E.W. on $\rm C_{31}$ follows as a consequence.

The secondary, marginal  correlation between $\rm H\alpha$ and $\rm C_{31}$ implies, however, that in bulge-dominated systems the $\rm H\alpha$ E.W. might
not provide a reliable estimate of the disk SFR, as argued by Devereux \& Young (1991). 
Since by definition the E.W. is computed
from the $\rm H\alpha$ net flux normalized to the red continuum, it would result artificially reduced by a strong contribution of the bulge to the continuum. \nl
To assess this important point we investigate the correlation between the $\rm H\alpha$ E.W.
and another SFR indicator, namely the $\rm H\alpha$ surface brightness ($\rm \Sigma~H\alpha$). The latter parameter
is defined as the ratio of the $\rm H\alpha$ net line flux to the disk area (computed using $\rm a_{25}$, the
diameter measured in the B band at the 25 \marc \ isophote). The normalizing
area is independent of the bulge properties, thus $\rm \Sigma~H\alpha$ should give a reliable estimate
of the disk SFR. In Fig. 8 we plot (adopting the usual symbols coded according to the
bulge-to-disk ratio) the two SFR parameters one against the other (panel a): the two quantities are linearly correlated
and disks and bulges are nicely segregated along the diagonal line. Furthermore we show
in Fig. 8b that the correlation between $\rm \Sigma~H\alpha$ and $\rm L_H$ is qualitatively similar 
to the one found between $\rm H\alpha$ E.W. and $\rm L_H$ (Fig. 5b). \nl
We conclude that $\rm H\alpha$ E.W. 
is marginally affected by the bulge contribution, thus it can be used as
a reliable estimate of the integrated SFR. 
Moreover another advantage of using $\rm H\alpha$ E.W.
over  $\rm \Sigma~H\alpha$ is that the accuracy of the former is about 10\%, while the uncertainty
on the area determination, up to 100\%, reflects into a similar uncertainty on $\rm \Sigma~H\alpha$.

\subsection{The Frequency Distribution of the $\rm H\alpha$ E.W.}

The analysis carried out in the previous section has shown that the SFR in disk galaxies 
is anti-correlated with their luminosity (mass). This finding makes the determination 
of the characteristic SFR properties of galaxies meaningless, unless a luminosity range is 
specified, i.e. if samples selected according to well defined completeness criteria are used. 
The $\rm H\alpha$ survey presented in this paper contains in fact a subsample complete to $\rm M_p = 
-18.8$, being selected from a magnitude limited subsample complete down to $\rm m_p=15.4$, entirely 
composed of objects lying at a constant distance (70 Mpc) (containing both isolated and cluster members).
With these data we construct the frequency distribution of the $\rm H\alpha$ E.W. in bins of Log 
E.W.=0.4, as given in Fig. 9a. It appears that, for galaxies brighter than $\rm M_p = 
-18.8$, the average $\rm H\alpha$ E.W. is 22 $\pm$ 2 $\rm \AA$. This value would increase if an intrinsically fainter sample was selected.
The distribution peaks at E.W.=25 $\rm \AA$. Over 40 \% of galaxies have 
their E.W. in the range 16-40 $\rm \AA$. Less than 15 \% galaxies have null E.W.: these
are either S0/S0a galaxies misclassified as spirals or high mass systems. 

\subsection{The Environmental Dependence of SFR}

Given the conclusions of the previous sections, it also follows that the comparison between the 
SFR properties of galaxies found in and outside clusters is meaningful only for samples with equal 
limiting optical luminosity. The frequency distributions for two different samples, one consisting of 40 cluster members, the other of 66 
isolated galaxies, both with a limiting magnitude $\rm m_p=15.4$, are compared in Fig. 9b and c. 
In Fig. 9b the membership is according to the criterion of Column 10 of Table 1 (``caustics''), 
while in Fig. 9c the membership is according to the criterion of Column 9 (angular separation).
As we can see from the latter figures, the distributions are insensitive to the adopted
membership criterion.
Moreover no significant difference is found between 
the cluster distribution and that for the isolated galaxies, as derived using the Kolmogorov-Smirnov test
(the probability that the cluster and isolated distributions derive from the same parent population is 2.5 \%).
This indicates, contrary to common 
believe, that cluster spiral+Irr galaxies have mean SFR values indistinguishable from those of the isolated objects. 
The histograms in Fig. 9 b and c suggest, however, that the cluster sample
has a less pronounced peak at Log $\rm H\alpha$ E.W. = 1.4 than the isolated one. 
Conversely there is a marginal evidence for an overabundance of intermediate $\rm H\alpha$ E.W.
(0-10 $\rm \AA$) among cluster objects. However, based upon the data in our possession, these differences are not statistically significant.

It is well known that late-type galaxies avoid the central regions of rich clusters. 
This is clearly the case in the Coma cluster (see Andreon 1996), and to a lesser extent in A1367.
It has been claimed that spirals at the periphery of the Coma cluster have bluer than
average color indices. Donas et al. 1995, for example, find an enhancement of objects with blue UV-B excess. 
If this reflects an enhanced SFR, it is expected that a similar pattern should show up
in our data. The radial distributions of the $\rm H\alpha$ E.W. are given up to 7 degrees
projected separation from Coma and A1367 
in Fig. 10 a and b respectively. Points are coded according to $\rm L_H$ to stress once more
the inverse proportionality between SFR and mass. 
Except for a marginal increase of the dispersion near the center of A1367, the only
apparent pattern is a relatively large fraction of faint (Log $\rm L_H~\leq$ 10.0) galaxies with large 
$\rm H\alpha$ E.W. in the shell contained between 1 and 2 degrees of projected radii around Coma.
However, we suggest that this effect is due to a mass segregation instead of a real enhancement of the SFR. 
In fact if we remove, to the first order, the luminosity (mass) dependence, by multiplying $\rm H\alpha$ E.W. by the corresponding $\rm L_H$ (as derived in three intervals in section 5.1), any evidence for radial gradients is canceled. (See Fig. 10 c and d).

\section{Discussion and Conclusions}

The main results of the present work can be summarized as follows: \nl
i) The present SFR properties of spiral galaxies, as derived from a representative sample of the local 
Universe, increase with increasing Hubble type.\nl
ii) These properties do not show a significant dependence on the explored range of local galaxy density,
spanning one order of magnitude between isolated supercluster objects and members of Coma-like 
clusters. \nl
iii) The present SFR shows a definite negative trend with the H band luminosity. The
average $\rm H\alpha$ E.W. increases from 0 to about 30 $\rm \AA$ with decreasing Log $\rm L_H$ from 11 to
9 $\rm L_{\sun}$, i.e. with decreasing dynamical mass from Log M = 11.5 to 9.5 $\rm M_{\sun}$.
This statement holds strictely for disk galaxies included in the present analysis. Nothing can
be concluded about the mass dependence in E+S0s.
$\rm H\alpha$ E.W. depends also marginally on the bulge-to-disk ratio, but this comes as
a consequence of a primary dependence of the bulge-to-disk ratio on $\rm L_H$. 

Point ii) does not mean that the environment is playing no role in
the formation and evolution of galaxies. The existence of a strong morphology-density relation, i.e. the 
environmental dependence of the fraction of early to late-type galaxies, is out of question. 
The conclusion of the present work is that
galaxies which retain a spiral-Irr morphology at the present cosmological epoch
have star formation properties which do not differ in and outside rich clusters.
Our observations do not rule out the possibility that galaxies in clusters form
primarily as E+SO and that spiral galaxies have only recently entered the dense cluster
environment, falling inward. If $\rm T_{res}$ is the time since their infall, $\rm \tau_{dep}$
is the HI depletion time-scale for ram-pressure (estimated to few $10^8$ yrs by Gavazzi 1989)
and $\rm \tau_{HI-H2}$ is the time-scale of the transformation from the atomic to the
molecular phase, then we can only conclude that $\rm \tau_{HI-H2}~>>~10^8$ yrs, i.e. that these
systems are still ``burning'' the molecular gas which was formed before they entered the
cluster environment. This is consistent with the evidence that the molecular gas content of HI deficient 
galaxies in rich clusters is normal (see e.g. Boselli et al. 1997).
It cannot be excluded, however, that on a time-scale longer than $\rm \tau_{HI-H2}$ these 
galaxies will run out of gas to fuel a substantial star formation rate, and hence they will progressively
evolve into anemic systems.  

The result of point iii) has deep implications on the models of galaxy evolution, following
the line traced by GPB96 and GS96 and the seminal work of Sandage (1986). GPB96 and GS96 argued that 
the correlation found between the B-V, U-B, B-H and UV-V color indices, other Population I 
indicators, and the galaxy mass follows 
from a basic dependence of galaxy evolution on the mass of their progenitor 
protogalaxies. \nl
Adopting a closed-box model (galaxies evolve out of the primeval gas, with 
some contribution from recycled gas) these authors claim that the presently observed color indices are consistent with the idea that galaxies are coeval systems (consistent with an age of 6.5-10 Gyrs), 
and that the time dependence of their initial collapse (SFR) follows an 
exponential with the time constant $\rm \tau$ inversely proportional to their mass, as $\rm \tau\propto M^{-3.1}$. 
Massive disk galaxies ($\rm M=10^{12.5}~M_{\sun}$) have a time constant as short as $5~10^8$ yrs, while low 
mass systems ($\rm M=10^{8.5}~M_{\sun}$) are consistent with a star formation history of $10^{10}$ yrs duration. \nl
Adapting the model of GS96 to an age of 10 Gyrs (consistently with KTC94) the dependence of $\rm \tau$ on mass: $\rm \tau\propto M^{-\alpha}$ requires $\rm \alpha$=2.5. 
Combining this key argument with the model of KTC94, which predicts 
the present SFR as a function of $\rm \tau$, we can derive SFR as a function of mass, as 
represented with the broken line in Fig. 5b, 6 and 7a. We use the model of KTC94 restricted to the case of an exponentially declining SFR with time (case b$\le$1), 
assuming a Salpeter IMF and solar metallicity (see their Table 1).
The model $\rm H\alpha$ E.W. are multiplied by 1.5
to account for the fact that the observed ones contain an equal contribution from the 
satellite [NII] lines (Kennicutt 1992). 
The model prediction and the data are in good agreement. The SFR predicted by the model, instead,
would fall short with respect to the observed values by a factor of 2 if
a Scalo instead of a Salpeter IMF was assumed, or if an extinction correction of about
1 mag, as discussed by KTC94, was applyed.
Altogether we claim that the observed $\rm H\alpha$ values are found in satisfactory agreement 
with those predicted by using the simple (perhaps simplistic) assumption that the time-scale of 
the single episode of star formation depends on the system mass as $\rm \tau\propto M^{-2.5}$,
without invoking recent bursts of star formation (case b$>$1).
In other words, the best known Pop I indicator, namely the present massive SFR,
is predicted by means of a simple evolutionary model
where the mass is the principal parameter governing the collapse time-scale of protogalaxies.
Massive galaxies had a short, intense burst of star formation shortly after their collapse, thus retaining little 
gas to fuel the present star formation. Low-mass systems underwent a much
longer, less spectacular episode of star formation, which is still transforming a significant 
fraction of gas into stars at the present cosmological epoch.
We emphasize that the above argument applies strictly to disk (spiral) galaxies.

Our evolutionary scenario has been confirmed by observations of faint galaxies at high redshift. Quoting Cowie et al. (1996):
``The more massive forming galaxies seen at $\rm z=1$ to 3 are identified
as earlier type spirals, whose star formation rates are initially
high and then decline rapidly at $\rm z < 1$, while for later type spirals
and smaller mass irregulars the star formation rates at $\rm z < 1$ are lower,
and the formation process persists to redshifts much closer to the present epoch''. \nl
Togheter with GS96 we stress that from the present observations there is no compelling evidence 
that galaxies are not coeval systems, i.e. the epoch of their formation is a function of their mass, 
but that the duration of their collapse is inversely proportional to their mass.

\acknowledgments

We wish to thank R. Kennicutt for providing us with data taken with the Steward 2.3m 
telescope and I. Randone who contributed to the data reduction. We are grateful to the TAC of the San Pedro 
Martir observatory for the generous time allocation. We warmly thank the 
telescope operators of the San Pedro Martir Observatory. 
We are grateful to an unknown referee whose comments helped improving the present work. 
This work has been partially supported by CONACYT research grant No. 211290-5-1430PE 
(L.C.) and by CARIPLO (G.G.).


\newpage

\figcaption[gavazzi.fig1.ps]{The sky distribution of CGCG galaxies which are members of 
the Coma Supercluster
($\rm 5000<V<10000~km~s^{-1}$). Dots represent E+S0 galaxies and empty circles spiral galaxies. Filled 
circles are spirals with an $\rm H\alpha$ measurement from the literature. Circled-filled symbols indicate 
galaxies observed in the present work. \label{fig1}}

\figcaption[gavazzi.fig2.ps]{The filter transmission profiles, (a) 1990-1995, (b) 1996-1997. \label{fig2}}

\figcaption[gavazzi.fig3a-o.ps]{Gray-scale representation of the off (top panels) and net (on-off, bottom 
panels) frames of all galaxies with a net $\rm H\alpha$ flux (a) and of 11 galaxies with null or 
negative net flux (b). The size of the displayed field is given in parenthesis. 
North is at the top and East to the left. \label{fig3}}

\figcaption[gavazzi.fig4.ps]{Comparison between the E.W. (a) and flux (b) measurements from 
this work and those found in the literature. \label{fig4}}


\figcaption[gavazzi.fig5.ps]{The correlation between $\rm H\alpha$ E.W. and morphological type (a) and H band luminosity (b).  
Types are given by discrete numerical classes. To avoid over-plotting a random number between 
-0.3 and 0.3 has been added to each numerical type.
Filled dots represent disk galaxies, empty circles represent bulge galaxies, crosses are used
for those objects whose structural information is not available. Galaxies with null $\rm H\alpha$ E.W. are
plotted at Log $\rm H\alpha E.W.= -0.1$.
The broken line gives the model adapted from KTC94 to contain the inverse dependence of $\rm \tau$ on 
mass, as discussed in Sec. 6. \label{fig5}}

\figcaption[gavazzi.fig6.ps]{The correlation between $\rm H\alpha$ E.W. and H band luminosity in bins of Hubble type (a, b, c). 
The same correlation is given separately for disk (d) and bulge (e) galaxies.
Same symbols as in Fig. 5.\label{fig6}}

\figcaption[gavazzi.fig7.ps]{The mutual relations between $\rm H\alpha$ E.W., $\rm L_H$ 
and $\rm C_{31}$. Same symbols as in Fig. 5.\label{fig7}}

\figcaption[gavazzi.fig8.ps]{The correlation between $\rm H\alpha$ E.W. and $\rm \Sigma~H\alpha$ (in $\rm erg~cm^{-2}~s^{-1}~arcmin^{-2}$) (a) and between $\rm \Sigma~H\alpha$ and $\rm L_H$ (b). Same symbols as in Fig. 5. A straight line with a slope of one is given in (a). \label{fig8}}

\figcaption[gavazzi.fig9.ps]{Frequency distribution of the $\rm H\alpha$ E.W. for the complete sample of isolated and cluster galaxies with $\rm M_p \leq -18.8$ (panel a). Isolated galaxies (full line) and cluster members according to the criterion given in Column 9 of Table 1 (dotted line) are given in panel b); Isolated galaxies and cluster members according to the criterion given in Column 10 of Table 1 are given in panel c). \label{fig9}}

\figcaption[gavazzi.fig10.ps]{Distributions of the $\rm H\alpha$ E.W. in the A1367 and Coma cluster (panels a, b) as a function of the projected angular separation from the X-ray centroids (in Degrees).
The vertical lines give the approx. extent of the Abell clusters (compare with Fig. 1).
Panels c) and d) give the projected radial distributions of the ``normalized'' $\rm H\alpha$ E.W.
This parameter helps removing the inverse dependence between the $\rm H\alpha$ E.W. and the
H band luminosity (see section 5.1). Squares mark galaxies with $\rm L_H < 10.0$; filled circles
mark objects with $\rm 10.0 < L_H < 10.5$; empty circles mark galaxies with $\rm L_H > 10.5$.\label{fig10}}

\end{document}